\documentclass[floatfix,twocolumn,showpacs,preprintnumbers,amsmath,amssymb,pra,superscriptaddress,longbibliography]{revtex4-1}
\usepackage{color}
\usepackage[usenames,dvipsnames,svgnames,table]{xcolor}
\usepackage[colorlinks=true,linkcolor=blue,urlcolor=blue,citecolor=blue]{hyperref}
\usepackage{mathtools}
\usepackage{graphicx}
\usepackage{dcolumn}
\usepackage{array}
\usepackage{lipsum}
\usepackage{bm}
\usepackage{subfigure}
\usepackage{amssymb}
\usepackage{multirow}
\usepackage{tabularx}
\usepackage{amsmath}
\usepackage{braket}
\usepackage{csquotes}
\graphicspath{{plots/}}
 \usepackage{lipsum}
\usepackage{mathrsfs}
\usepackage{MnSymbol}

\newcommand{\beq}{\begin{equation}}
\newcommand{\eeq}{\end{equation}}
\newcommand{\bea}{\begin{eqnarray}}
\newcommand{\eea}{\end{eqnarray}}

\begin{document}
\title{Direct free energy calculation from \emph{ab initio} path integral Monte Carlo simulations of warm dense matter}

\author{Tobias Dornheim}
\email{t.dornheim@hzdr.de}

\affiliation{Center for Advanced Systems Understanding (CASUS), D-02826 G\"orlitz, Germany}
\affiliation{Helmholtz-Zentrum Dresden-Rossendorf (HZDR), D-01328 Dresden, Germany}

\author{Zhandos~A.~Moldabekov}

\affiliation{Center for Advanced Systems Understanding (CASUS), D-02826 G\"orlitz, Germany}
\affiliation{Helmholtz-Zentrum Dresden-Rossendorf (HZDR), D-01328 Dresden, Germany}

\author{Sebastian Schwalbe}

\affiliation{Center for Advanced Systems Understanding (CASUS), D-02826 G\"orlitz, Germany}
\affiliation{Helmholtz-Zentrum Dresden-Rossendorf (HZDR), D-01328 Dresden, Germany}

\author{Jan Vorberger}
\affiliation{Helmholtz-Zentrum Dresden-Rossendorf (HZDR), D-01328 Dresden, Germany}

\begin{abstract}
We carry out highly accurate \emph{ab initio} path integral Monte Carlo (PIMC) simulations to directly estimate the free energy of various warm dense matter systems including the uniform electron gas and hydrogen without any nodal restrictions or other approximations. Since our approach is based on an effective ensemble in a bosonic configuration space, it does not increase the computational complexity beyond the usual fermion sign problem. Its application to inhomogeneous cases such as an electronic system in a fixed external ion potential is straightforward and opens up the enticing possibility to benchmark density functional theory and other existing methods. Finally, it is not limited to warm dense matter, and can be applied to a gamut of other systems such ultracold atoms and electrons in quantum dots.
\end{abstract}

\maketitle


The rigorous theoretical description of \emph{warm dense matter} (WDM)---an extreme state that naturally occurs in astrophysical objects such as giant planet interiors~\cite{Benuzzi_Mounaix_2014} and white dwarfs~\cite{Kritcher2020}, and which is relevant to cutting-edge technological applications such as inertial fusion energy~\cite{AbuShawareb_PRL_2024,Hurricane_RevModPhys_2023,hu_ICF}---constitutes one of the most pressing challenges in a variety of fields~\cite{wdm_book,new_POP,Dornheim_review,bonitz2024principles} including plasma physics, material science, and quantum chemistry.
In the WDM regime, the Wigner-Seitz radius $r_s=d/a_\textnormal{B}$, the degeneracy temperature $\Theta=k_\textnormal{B}T/E_\textnormal{F}$ (where $E_\textnormal{F}$ is the Fermi energy~\cite{quantum_theory}), and the coupling parameter $\Gamma=W/K$ (where $W$ and $K$ are the interaction and kinetic energy, respectively) are all of the order of unity~\cite{Ott2018}, implying a complex interplay of effects such as Coulomb coupling, 
quantum degeneracy and delocalization, strong thermal excitations, and partial ionization.
Accurate simulations of WDM thus require a simultaneous and holistic treatment of these effects, which is notoriously challenging.

In this situation, the combination of thermal density functional theory (DFT)~\cite{Mermin_DFT_1965} with molecular dynamics (MD) simulations~\cite{wdm_book} has emerged as a widely used tool~\cite{Holst_PRB_2008,Witte_PRL_2017,kushal}, balancing an often acceptable level of accuracy with a manageable computational effort.
In principle, DFT would be exact if the exact exchange--correlation (XC) functional was provided as an external input.
In practice, however, the exact XC-functional is generally unknown except for a few simplified models. The accuracy of a given DFT simulation is thus decisively determined by the employed approximate functional, where different levels of sophistication are conveniently classified on Jacob's ladder of functional approximations~\cite{Perdew_AIP_2001}.

At ambient conditions, where the electrons are in their respective ground state, the accuracy of different functionals is reasonably well understood based on extensive comparisons to exact reference data e.g.~from quantum Monte Carlo simulations or from experiments~\cite{Goerigk_PCCP_2017}.
In the WDM regime, the situation is considerably less clear. First, thermal DFT simulations require a parametrization of the XC-free energy $f_\textnormal{xc}$ that explicitly depends on the temperature~\cite{kushal,karasiev_importance,Sjostrom_PRB_2014,bonitz2024principles}; the existing zoo of ground-state functionals~\cite{Goerigk_PCCP_2017} and their known performance in the limit of $T=0$ is thus of limited value. Second, the estimation of the free energy based on highly accurate quantum Monte Carlo simulations is substantially more complicated than the estimation of the XC-energy in the ground state; the situation is particularly dire for inhomogeneous systems such as a system of electrons in a fixed set of nuclei coordinates---the standard problem of thermal DFT---for which even the evaluation of the adiabatic connection formula is highly impractical, leading, to our knowledge, to an almost complete absence of reference data for $f_\textnormal{xc}$ for real systems. Third, experiments with WDM~\cite{falk_wdm} are notoriously hard to diagnose~\cite{Kasim_POP_2019,boehme2023evidence,Sperling_PRL_2015,Dornheim_T_2022,Tilo_Nature_2023}, which is further exacerbated by possible inhomogeneity~\cite{Chapman_POP_2014}, non-equilibrium~\cite{vorberger2023revealing} and other complications; their value for the rigorous benchmarking of existing XC-functionals has thus remained limited.

\begin{figure}\centering
\includegraphics[width=0.37\textwidth]{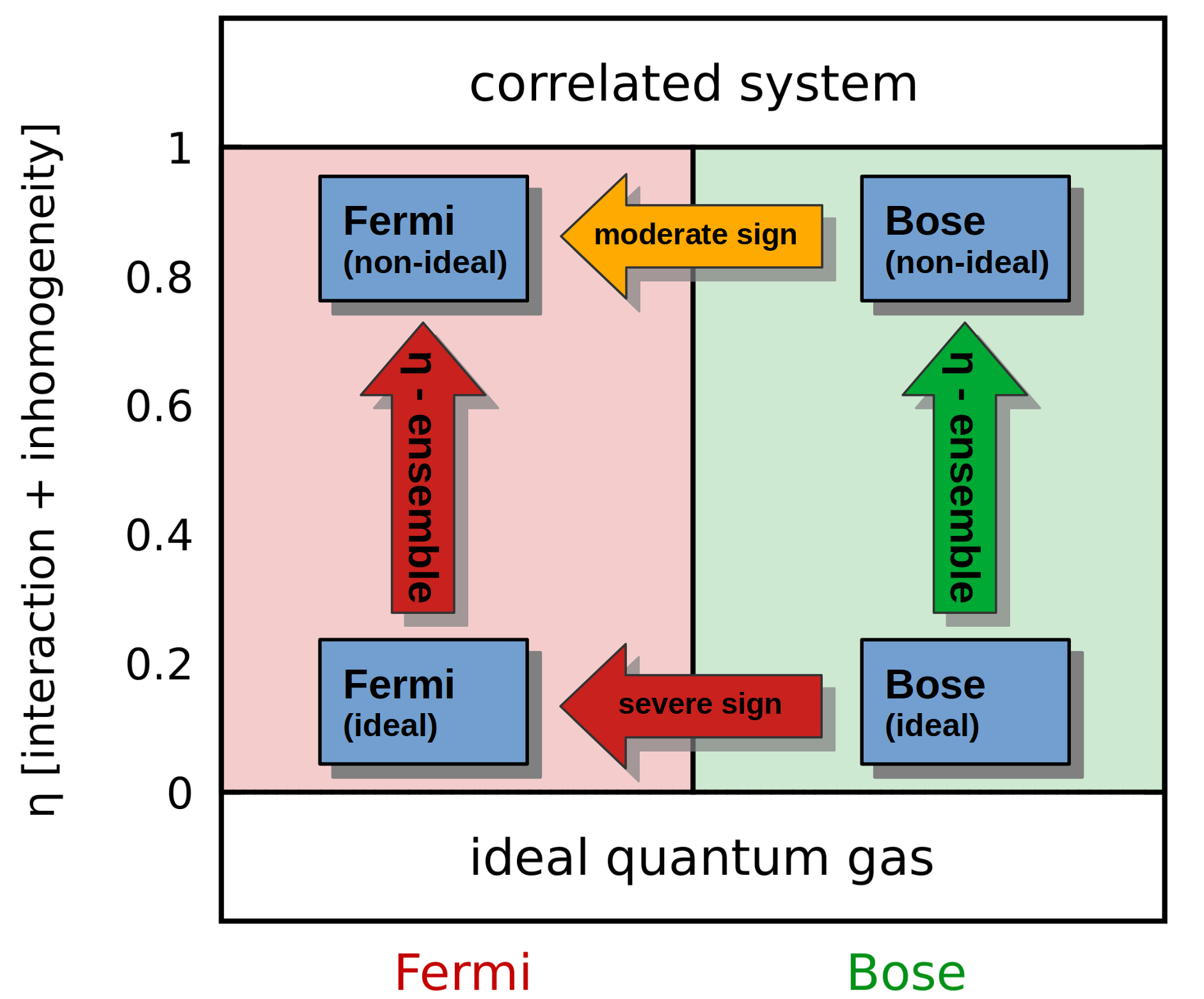}
\caption{\label{fig:scheme} Schematic illustration of our idea: 
using the extended ensemble approach, we map the free energy of the uniform, ideal Bose gas (bottom right) onto the interacting, inhomogeneous Bose system (top right) in the sign-problem free sector (green); as the final step, we recover the correct fermionic result of interest (top left) from a fermionic PIMC simulation, where the sign problem is comparably moderate; this makes it strongly preferable to the alternative route in the fermionic sector, where the sign problem becomes substantially more severe in the limit of $\eta\to0$ (bottom left).
}
\end{figure}

In this Letter, we change this unsatisfactory situation by implementing a new approach for the direct estimation of the free energy of arbitrary equilibrium systems based on exact \emph{ab initio} path integral Monte Carlo (PIMC) simulations~\cite{cep,review,Bohme_PRE_2023,Dornheim_Science_2024}. This method does not increase the computational complexity beyond the usual fermion sign problem~\cite{dornheim_sign_problem,troyer} of the interacting system of interest and it can be readily applied even to inhomogeneous systems without the need for a cumbersome numerical inversion for an adiabatic connection formula for which a density has to be kept constant for different coupling strengths~\cite{Wagner_PRB_2014}.
To demonstrate the versatility of our idea, we apply it to three representative cases: i) we benchmark our implementation of the proposed extended ensemble approach (cf.~Fig.~\ref{fig:scheme}) for a system of harmonically perturbed fermions against the density stiffness theorem in the linear response regime~\cite{moldabekov2024density,quantum_theory}; ii) we directly compute $f_\textnormal{xc}$ of the warm dense electron gas and find excellent agreement with the existing parametrization by Groth \emph{et al.}~\cite{groth_prl} that has been obtained based on a thermodynamic integration over the density parameter; iii) as the capstone of our work, we estimate the electronic free energy within a fixed proton configuration and compare with thermal DFT simulations using both thermal and ground-state XC-functionals.

We are convinced that our idea will be of high value for the study of warm dense matter by providing a systematic pathway towards the categorization of XC-functionals; this might be particularly valuable for the development of new functionals that are specifically designed for the application at finite temperatures~\cite{ksdt,groth_prl,Karasiev_PRL_2018,Karasiev_PRB_2022,kozlowski2023generalized,pribram}.
Moreover, having access to the free energy will allow one to check existing equation-of-state databases~\cite{Militzer_PRE_2021} that have been constructed based on other observables (e.g.~internal energy and pressure) for thermodynamic consistency. Finally, we note that our approach is not limited to WDM and can be readily applied to a gamut of other systems including ultracold atoms~\cite{cep,Dornheim_SciRep_2022,Filinov_PRA_2012} and electrons in quantum dots~\cite{Reimann_RevModPhys_2002,Dornheim_NJP_2022}.

\textbf{Idea.} The Hamiltonian of $N$ unpolarized electrons in an external potential $v_\textnormal{ext}(\mathbf{r})$ is given by
\begin{eqnarray}\label{eq:Hamiltonian}
    \hat{H} = -\frac{\hbar^2}{2m_e}\sum_{l=1}^N \nabla_l^2 + \eta\left\{ \sum_{l=1}^N v_\textnormal{ext}(\hat{\mathbf{r}}_l) + \sum_{l>k}^N W_\textnormal{E}(\hat{\mathbf{r}}_k,\hat{\mathbf{r}}_l)\right\}\ ,
\end{eqnarray}
where $W_\textnormal{E}(\mathbf{r},\mathbf{s})$ is the usual Ewald potential as it has been introduced e.g.~in Ref.~\cite{Fraser_PRB_1996}.
Here $\eta\in[0,1]$ is a free parameter that will be specified below; the physical Hamiltonian that describes a given system of interest is recovered for $\eta=1$.
The basic idea of the \emph{ab initio} PIMC method is to stochastically sample the thermal density matrix $\hat\rho=e^{-\beta\hat{H}}$ in coordinate representation, where $\beta=1/k_\textnormal{B}T$ is the inverse temperature. This gives one, in principle, access to all thermodynamic properties of the system except for the partition function $Z=\textnormal{Tr}\hat{\rho}$ itself; direct access to the free energy energy $F=-\beta^{-1}\textnormal{log}(Z)$ is thus precluded.

On the other hand, it is well-known that PIMC simulations give one access to the ratio of partition functions or, equivalently, free energy differences between two or more systems~\cite{Lyubartsev_JCP_1992,boninsegni2,Dornheim_PRB_nk_2021}.
A particularly infamous case is given by the \emph{average sign} in direct PIMC simulations (i.e., without anti-symmetrized imaginary-time propagators)~\cite{dornheim_sign_problem}
\begin{eqnarray}\label{eq:sign}
    S = \frac{1}{Z_\textnormal{Bose}} \sumint \textnormal{d}\mathbf{X}\ |W(\mathbf{X})|\ \textnormal{sgn}\left( W(\mathbf{X}) \right) = \frac{Z_\textnormal{Fermi}}{Z_\textnormal{Bose}}\ ,
\end{eqnarray}
where $\sumint \textnormal{d}\mathbf{X}$ indicates the combined integration over all possible paths $\mathbf{X}$ and sum over all possible permutations of particle coordinates, see Ref.~\cite{Dornheim_permutation_cycles} for a more extensive discussion; $W(\mathbf{X})$ denotes the fermionic configuration weight that can be both positive and negative, and its modulus corresponds to the corresponding Bose system.
Hence, one automatically estimates the free energy difference between fermions and bosons in any given signful direct PIMC simulation,
\begin{eqnarray}\label{eq:delta_F_sign}
    F_\textnormal{Fermi} - F_\textnormal{Bose} = - \frac{1}{\beta} \textnormal{log}\left( S \right)\ .
\end{eqnarray}
By itself, however, Eq.~(\ref{eq:delta_F_sign}) is of rather limited practical value, as the free energy of a non-ideal Bose system $F_\textnormal{Bose}$ is generally also unknown.

To overcome this obstacle, we generalize the spirit of Eq.~(\ref{eq:delta_F_sign}) to the estimation of a pair of, in principle, arbitrary systems "$a$" and "$b$"; the ratio of the corresponding partition functions can be estimated from the generalized ensemble 
    $Z_\textnormal{extended} = c_\eta Z_a + Z_b$
via the ratio
\begin{eqnarray}\label{eq:ratio}
    r_{a,b} = {c_\eta} \frac{Z_a}{Z_b} =  \frac{\braket{\hat{\delta}_a}_\textnormal{extended}}{\braket{\hat{\delta}_b}_\textnormal{extended}}\ ,
\end{eqnarray}
with $\hat{\delta}_a$ and $\hat{\delta}_b$ only counting configurations from their respective configuration space, and $c_\eta$ being an arbitrary constant that can be chosen freely to optimize ergodicity~\cite{boninsegni2}.
Both systems are governed by the Hamiltonian Eq.~(\ref{eq:Hamiltonian}), with "$a$" and "$b$" corresponding to $\eta=1$ and $\eta=0$, respectively. The free energy of an ideal, uniform $N$-particle Bose gas $F_{\textnormal{Bose},\eta=0}$ can be easily computed from a recursion relation of $Z$~\cite{krauth2006statistical,DuBois}, see Ref.~\cite{supplement} for details.
Combining Eqs.~(\ref{eq:delta_F_sign}) and (\ref{eq:ratio}), the full fermionic free energy of interest is given by
\begin{eqnarray}\label{eq:final}
    F_\textnormal{Fermi} = F_{\textnormal{Bose},\eta=0} - \frac{1}{\beta}\left\{
\textnormal{log}\left(\frac{r_{a,b}}{c_\eta}\right) + \textnormal{log}\left(S\right)
    \right\}\ .
\end{eqnarray}
The full idea is summarized in Fig.~\ref{fig:scheme}. Using the $\eta=0$ limit (bottom right) as a known reference, we estimate the free energy of the interacting and inhomogeneous Bose system $F_\textnormal{Bose}$ from the $\eta$-ensemble, see the green arrow. This step can be decomposed into an arbitrary number of intermediate steps $\eta_i\in(0,1)$, see Ref.~\cite{supplement}, and is carried out in the sign-problem free bosonic sector where the computation cost is low. Finally, the free energy of the fermionic WDM system of interest is estimated from the average sign $S$ (orange arrow); since the latter is computed for the interacting system, the sign problem is relatively moderate. This is in stark contrast to alternative routes that would require fermionic PIMC simulations in the ideal limit (red arrows in Fig.~\ref{fig:scheme}).


\textbf{Results.} To illustrate our approach, and to rigorously benchmark its implementation into the \texttt{Ishtar} code~\cite{ISHTAR}, we consider an extended ensemble consisting of "b" a uniform ideal Fermi gas with $N=14$, $r_s=2$, and $\Theta=4$ and "a" the same system subject to an external harmonic perturbation of wave vector $\mathbf{q}$ and perturbation amplitude $A$~\cite{moroni,moroni2,Dornheim_PRL_2020,Bohme_PRL_2022}. In the linear-response limit of $A\to0$, the corresponding ratio of the partition functions follows from the static linear density response function $\chi(\mathbf{q},0)$ via the density stiffness theorem~\cite{quantum_theory,moldabekov2024density}, see the Supplemental Material~\cite{supplement} for additional details.

\begin{figure}\centering
\includegraphics[width=0.45\textwidth]{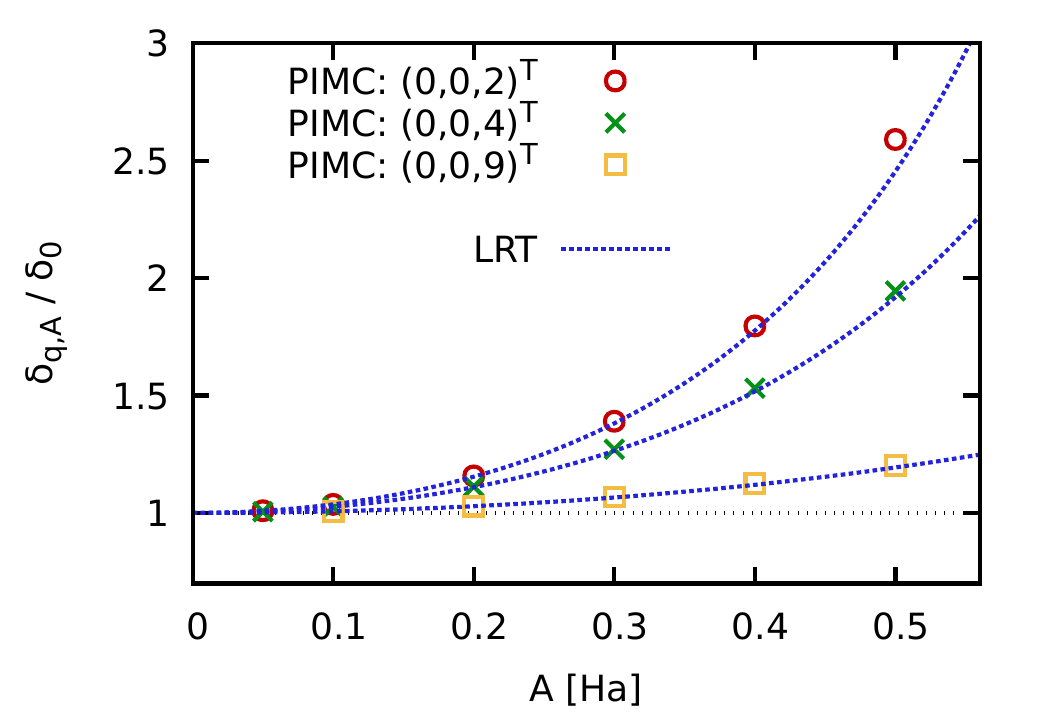}\\
\caption{\label{fig:stiffness} Ratio of the harmonically perturbed partition function $Z_{\mathbf{q},A}$ and the unperturbed partition function $Z_0$ for an ideal Fermi gas at $r_s=2$ and $\Theta=4$ with $N=14$. Symbols: PIMC results for different $\mathbf{q}$ in units of $2\pi/L$, where $L$ is the box length; dotted green: linear-response theory (LRT) limit computed from the density stiffness theorem~\cite{quantum_theory,moldabekov2024density,supplement}.
}
\end{figure}

In Fig.~\ref{fig:stiffness}, we show PIMC results for the fermionic ratio of partition functions for three different wave vectors as the red circles, green crosses, and yellow squares.
The dotted blue lines show corresponding predictions from linear-response theory, which are in excellent agreement with our simulation results for small $A$, as it is expected.
In addition to validating our approach and its implementation, these results also illustrate the possibility to compute linear (and potentially also nonlinear~\cite{Dornheim_JCP_2023}) response functions from simulation results for the free energy at finite temperatures; this might be more convenient than previous schemes based on the density in coordinate space~\cite{Moldabekov_JCTC_2022}, and could additionally be used to check consistency with the latter.

\begin{figure}\centering
\includegraphics[width=0.45\textwidth]{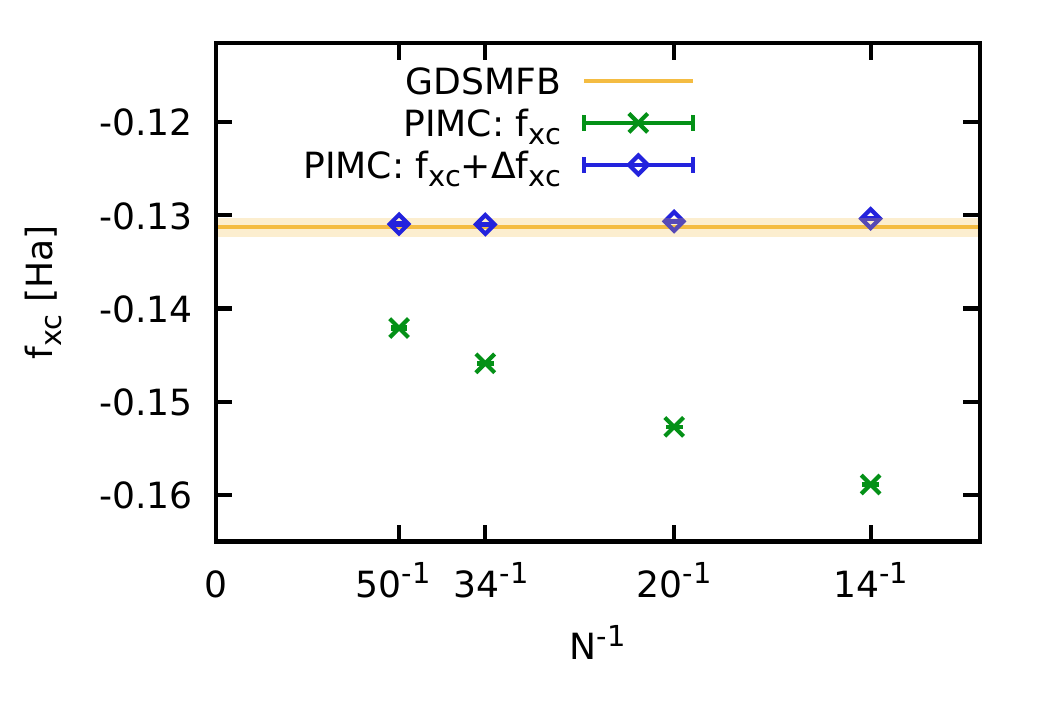}
\caption{\label{fig:UEG_rs3p23_theta2} XC-free energy per particle of the uniform electron gas at $r_s=3.23$ and $\Theta=2$ as a function of system size $N$. 
Green crosses: direct PIMC results [Eq.~(\ref{eq:final})]; blue diamonds: green crosses + finite-size correction $\Delta f_{xc}$, see Refs.~\cite{Malone,supplement}; yellow line: adiabatic-connection based previous parametrization by Groth \emph{et al.}~\cite{groth_prl}, where the shaded yellow area indicates an interval of $\pm1\,$mHa.
}
\end{figure}

As a second application, we compute the free energy of the uniform electron gas~\cite{review} at $r_s=3.23$ and $\Theta=2$. This density can be realized experimentally e.g.~in hydrogen jets~\cite{Zastrau,Fletcher_Frontiers_2022}, where it might give rise to exotic, hitherto unexplored phenomena such as the recently predicted roton-type feature in the dynamic structure factor~\cite{Hamann_PRR_2023}.
In Fig.~\ref{fig:UEG_rs3p23_theta2}, we show the XC-free energy (per particle) $f_\textnormal{xc}=(F-F_0)/N$
, where $F_0$ denotes the free energy of the ideal Fermi gas.
We note that $f_\textnormal{xc}$ is a key property for a host of applications including astrophysical models~\cite{Pothekin,SAUMON20221}, and allows for thermal DFT simulations on the level of the local density approximation~\cite{Sjostrom_PRB_2014,ksdt,groth_prl,karasiev_importance,kushal}.
The green crosses show our new PIMC results [Eq.~(\ref{eq:final})] for different system size $N$; we find a moderate dependence on the system size of the order of $\sim10\%$, which is consistent with previous studies of related properties~\cite{dornheim_prl,review}.
The blue diamonds have been obtained by adding to the raw PIMC results a finite-size correction $\Delta f_\textnormal{xc}$, which has been computed using the \texttt{uegpy} package by F.D.~Malone~\cite{uegpy}; see the Supplemental Material~\cite{supplement} for additional details.
The thus corrected results exhibit a nearly negligible residual dependence on $N$ and are within $\pm1\,$mHa (shaded yellow area) to the parametrization of $f_\textnormal{xc}$ by Groth \emph{et al.}~\cite{groth_prl} (yellow line).
We stress that the GDSMFB results have been obtained based on extensive results for the interaction energy covering a broad range of densities and temperatures via the adiabatic connection formula, whereas PIMC simulations for a single density--temperature combination have been sufficient for the present study.

\begin{figure}\centering
\includegraphics[width=0.45\textwidth]{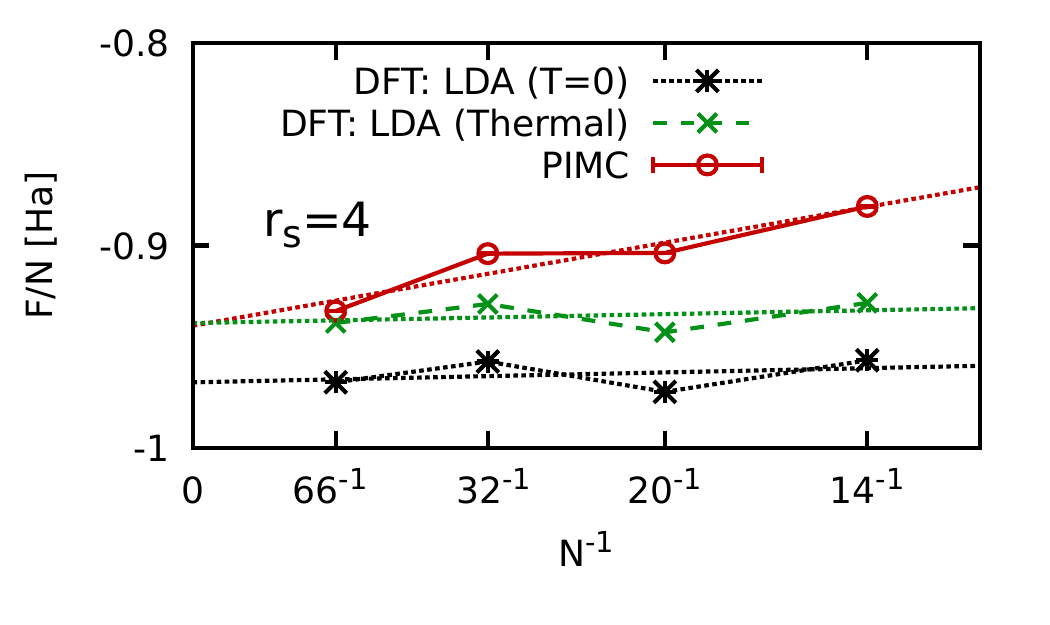}\\\vspace*{-1.25cm}
\includegraphics[width=0.45\textwidth]{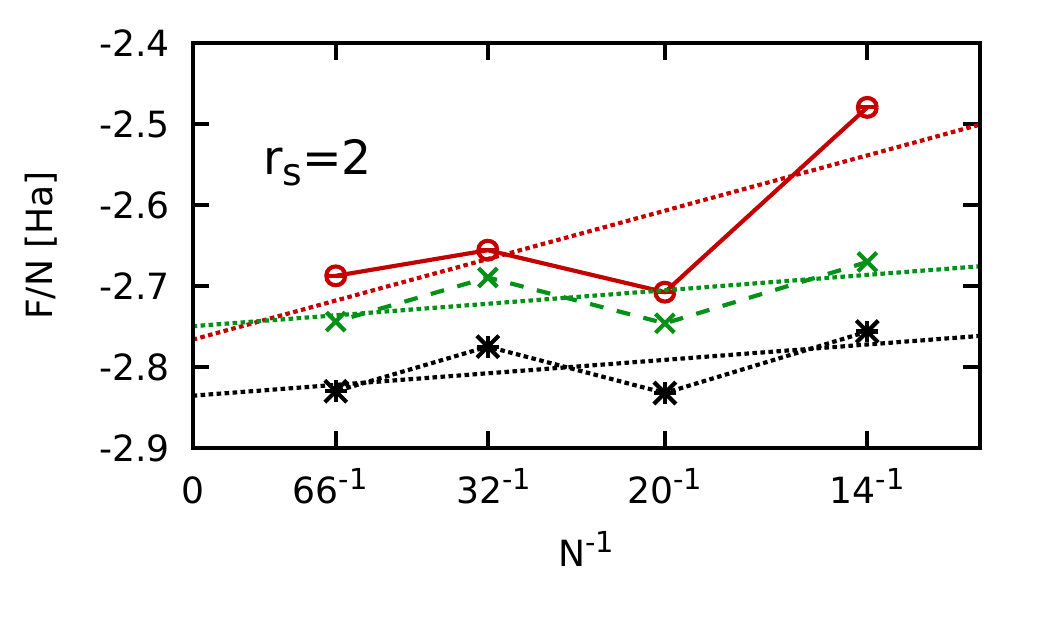}\vspace*{-0.5cm}
\caption{\label{fig:snap} Electronic free energy per electron of different hydrogen snapshots at $\Theta=2$ and $r_s=4$ [top] and $r_s=2$ [bottom] for different $N$. Red circles: PIMC; green crosses: DFT with thermal LDA functional~\cite{groth_prl}; black stars: DFT with ground-state LDA functional~\cite{Perdew_Wang_PRB_1992}. The dotted lines correspond to empirical linear fits.
}
\end{figure}

As the final example, we compute the electronic free energy in the fixed external potential of a proton configuration at $\Theta=2$ and two relevant densities ($r_s=4$ and $r_s=2$) for different numbers of hydrogen atoms in Fig.~\ref{fig:snap}. The red circles show our new PIMC results, and the green crosses and black stars show DFT results for the same snapshots using the thermal LDA XC-functional by Groth \emph{et al.}~\cite{groth_prl} and the ground-state XC-functional by Perdew and Wang~\cite{Perdew_Wang_PRB_1992}, respectively.
First, we note that all data sets exhibit similar fluctuations with $N$, which are a consequence of the specific proton configurations in the respective snapshots.
In addition, there appears a more systematic dependence on $N$ that is somewhat different between PIMC and DFT, see the dotted lines showing empirical linear fits to the respective data.
Most importantly, we find that consistently taking into account thermal effects in the XC-functional leads to a substantially improved agreement towards PIMC both for finite $N$ and in the thermodynamic limit.

A more detailed comparative analysis of finite-size effects in PIMC and DFT calculations, and a systematic study of the impact of inhomogeneity effects onto the level of accuracy of different XC-functionals 
that takes into account more sophisticated functionals~\cite{Karasiev_PRL_2018,Karasiev_PRB_2022,kozlowski2023generalized} for different densities and temperatures (and potentially different light elements) 
will be pursued in dedicated future investigations.

\textbf{Discussion.} We have implemented a new approach for the direct computation of the free energy of interacting quantum many-body systems such as warm dense matter from \emph{ab initio} PIMC simulations. Being based on an extended ensemble in the configuration space of bosons, this approach basically comes at no additional computation cost compared to the usual fermion sign problem in equilibrium PIMC simulations~\cite{dornheim_sign_problem}.
To demonstrate the versatility of our idea, we have applied it to three representative practical examples. First, we have simulated a harmonically perturbed ideal Fermi gas, where the ratio of the partition functions is governed by the density stiffness theorem in the linear response regime. We have found perfect agreement between our simulation results and the theoretical prediction. In addition to their value as a validation, these results illustrate the possibility to estimate a variety of linear and nonlinear~\cite{Dornheim_JCP_2023} response properties based solely on simulation results for the free energy, which might be particularly relevant for thermal DFT simulations~\cite{Moldabekov_JCTC_2022}.
Second, we have directly computed the XC-free energy of the warm dense uniform electron gas. Our new results are in excellent agreement with the existing parametrization by Groth \emph{et al.}~\cite{groth_prl} that is based on the thermodynamic integration over an extensive set of state points. In stark contrast, our scheme allows to estimate $f_\textnormal{xc}$ just from PIMC simulations at a single density-temperature point.
A particular strength of our approach is that it straightforwardly allows for the estimation of the free energy of inhomogeneous systems without the difficult requirement of keeping the density constant for an integration along an adiabatic connection; the third application presented in this Letter is thus the free energy of a system of electrons in the external potential of a fixed ion configuration, which is the standard problem in thermal DFT-MD simulations.
The comparison of our PIMC reference data with DFT results has 
nicely illustrated the importance of thermal XC-effects for the estimation of the free energy, with the thermal LDA being fairly accurate (we find systematic errors of $\sim1\%$ here) at the present conditions. We note that all PIMC results are available in an online repository~\cite{repo}.

We are convinced that our approach opens up a multitude of new avenues for interesting and impactful future research. The estimation of the free energy, and in this way also of the partition function $Z$ and of the entropy $\sigma$, is interesting in its own right and will give new insights into the physics of interacting quantum many-body systems. Indeed, the proposed scheme is not specific to the study of WDM and can be easily generalized to PIMC simulations of a host of other systems such as ultracold atoms~\cite{cep,Dornheim_SciRep_2022,Filinov_PRA_2012} and electrons in quantum dots~\cite{Reimann_RevModPhys_2002,Dornheim_NJP_2022}.
A particularly interesting application is given by the estimation of the free energy of real warm dense matter systems such as hydrogen~\cite{dornheim2024ab,Dornheim_JCP_2024,bonitz2024principles} and beryllium~\cite{Dornheim_Science_2024}, and potentially light mixtures such as lithium hydride~\cite{LiH_PNAS_2009}. In this regard, the direct access to $F$ might help to test the thermodynamic consistency of existing equation-of-state tables that are based on other observables such as the pressure.
Arguably the most important use case of the presented PIMC approach to the free energy is to rigorously benchmark XC-functionals for DFT simulations, and to guide the development of new functionals that explicitly take into account temperature effects~\cite{ksdt,groth_prl,Sjostrom_PRB_2014,Karasiev_PRL_2018,Karasiev_PRB_2022,kozlowski2023generalized}. Here, the availability of exact reference data has the potential to be a true game changer, and could facilitate the inversion of the exact XC-functional~\cite{Kasim_PRL_2021}, XC-potential, or thermal enhancement factors in selected cases.
Finally, we mention the important possibility to combine the present idea with existing methodologies that alleviate the fermion sign problem~\cite{Dornheim_NJP_2015,Hirshberg_JCP_2020,Dornheim_JCP_2020,Xiong_JCP_2022,Dornheim_JCP_xi_2023,Dornheim_JPCL_2024,filinov2023equation}, which might give access to a broader region of the parameter space that is of interest for contemporary WDM research and beyond.


\begin{acknowledgements}
We acknowledge stimulating discussions with Paul E.~Grabowski, and valuable feedback by Panagiotis Tolias.

This work was partially supported by the Center for Advanced Systems Understanding (CASUS), financed by Germany’s Federal Ministry of Education and Research (BMBF) and the Saxon state government out of the State budget approved by the Saxon State Parliament. This work has received funding from the European Research Council (ERC) under the European Union’s Horizon 2022 research and innovation programme
(Grant agreement No. 101076233, "PREXTREME"). 
Views and opinions expressed are however those of the authors only and do not necessarily reflect those of the European Union or the European Research Council Executive Agency. Neither the European Union nor the granting authority can be held responsible for them. Computations were performed on a Bull Cluster at the Center for Information Services and High-Performance Computing (ZIH) at Technische Universit\"at Dresden, at the Norddeutscher Verbund f\"ur Hoch- und H\"ochstleistungsrechnen (HLRN) under grant mvp00024, and on the HoreKa supercomputer funded by the Ministry of Science, Research and the Arts Baden-W\"urttemberg and
by the Federal Ministry of Education and Research.
\end{acknowledgements}

\bibliography{bibliography}

\begin{thebibliography}{82}
\bibitem [{rep()}]{repo}%
  \BibitemOpen
  \href@noop {} {}\bibinfo {note} {A link to a repository containing all PIMC results will be made available upon publication.}\BibitemShut {Stop}%
\end{thebibliography}
\end{document}


\title{ 
{Supplemental Material:}
Direct free energy calculation from \emph{ab initio} path integral Monte Carlo simulations of warm dense matter
}

\author{Tobias Dornheim}
\email{t.dornheim@hzdr.de}

\affiliation{Center for Advanced Systems Understanding (CASUS), D-02826 G\"orlitz, Germany}
\affiliation{Helmholtz-Zentrum Dresden-Rossendorf (HZDR), D-01328 Dresden, Germany}

\author{Zhandos~A.~Moldabekov}

\affiliation{Center for Advanced Systems Understanding (CASUS), D-02826 G\"orlitz, Germany}
\affiliation{Helmholtz-Zentrum Dresden-Rossendorf (HZDR), D-01328 Dresden, Germany}

\author{Sebastian Schwalbe}

\affiliation{Center for Advanced Systems Understanding (CASUS), D-02826 G\"orlitz, Germany}
\affiliation{Helmholtz-Zentrum Dresden-Rossendorf (HZDR), D-01328 Dresden, Germany}

\author{Jan Vorberger}
\affiliation{Helmholtz-Zentrum Dresden-Rossendorf (HZDR), D-01328 Dresden, Germany}

\maketitle



\section{Generalization to $M_\eta$ intermediate $\eta$-steps}\label{sec:intermediate}

The successful evaluation of Eq.~(4) in the main text with "$a$"$\to\eta=1$ and "$b$"$\to\eta=0$ presupposes a sufficient overlap between the respective configuration spaces in the PIMC simulation. In practice, this holds for weakly interacting systems in a weak external potential, but breaks down for strong coupling and/or strong inhomogeneity. The constant $c_\eta$ does not fundamentally change this balance; its function is mostly to cancel a constant energy offset, such as the interaction of the electrons with the homogeneous positive background in the uniform electron gas.

To apply the present idea to arbitrary systems, we propose to introduce $M_\eta$ intermediate $\eta$-steps, with the original idea being recovered for $M_\eta=0$. 
This leads to $M_\eta+1$ extended partition sums
\begin{eqnarray}
    Z^i_\textnormal{extended} = c_{\eta^i} Z_{\eta^i} + Z_{\eta^{i-1}}\ ,
\end{eqnarray}
and the corresponding ratios
\begin{eqnarray}
    r_{i,i-1} = \frac{\delta_{\eta^i}}{\delta_{\eta^{i-1}}} = c_{\eta^i} \frac{Z_{\eta^i}}{Z_{\eta^{i-1}}}\ ,
\end{eqnarray}
with $i=1,\dots,M_\eta+1$. The difference in free energy between the limits of $\eta=1$ and $\eta=0$ is given by
\begin{eqnarray}
    F_{\eta=1}-F_{\eta=0} = - \frac{1}{\beta} \sum_{i=1}^{M_\eta+1}\left\{
    \textnormal{log}\left( \frac{r_{i,i-1}}{c_{\eta^i}} \right)
    \right\}\ .
\end{eqnarray}
By increasing $M_\eta$, the difference between the configuration spaces of adjacent $\eta^i$ can be made arbitrarily small, thereby ensuring sufficient overlap. At the same time, this, too, does generally not lead to an increase in the computational complexity due to the absence of a sign problem for bosons along $\eta\in[0,1]$.

\section{Free energy of ideal uniform Bose reference system\label{sec:Jan}}

The partition function of $N^\uparrow$ spin-polarized ideal bosons in a volume $\Omega=L^3$ at inverse temperature $\beta$ can be computed from the recursion relation~\cite{krauth2006statistical}
\begin{eqnarray}\label{eq:recursion}
    Z_{N^\uparrow,\Omega}(\beta) = \frac{1}{N^\uparrow} \sum_{q=1}^{N^\uparrow} Z_{N^\uparrow-q}(\beta)Z_1(q\beta)\ ,
\end{eqnarray}
with $Z_{0,\Omega}(\beta)\equiv1$.
For an unpolarized Bose gas with $N=2N^\uparrow$ in the same volume and at the same inverse temperature, we simply have
\begin{eqnarray}\label{eq:Z_unpolarized_simple}
    Z_{N,\Omega}(\beta) = \left\{ Z_{N^\uparrow,\Omega}(\beta) \right\}^2 \ ,
\end{eqnarray}
and the corresponding total free energy follows from the relation $F=-\beta^{-1}\textnormal{log}(Z)$.


The final ingredient to evaluate Eqs.~(\ref{eq:recursion}) and (\ref{eq:Z_unpolarized_simple}) is given by the single-particle partition function, which, for a particle in a cubic box of length $L$ featuring periodic boundary conditions is given by~\cite{DuBois,Dornheim_permutation_cycles}
\begin{eqnarray}
    Z_{1,\Omega}(\beta) = \left\{  \sum_{x=-\infty}^\infty e^{-\beta E_x} \right\}^3 \ ,
\end{eqnarray}
with the energy [assuming Hartree atomic units]
\begin{eqnarray}
    E_x = \frac{1}{2} \left( \frac{2\pi x}{L} \right)^2 \ .
\end{eqnarray}

\section{Density stiffness theorem\label{sec:stiffness}}

To rigorously benchmark our implementation, we utilize the density stiffness theorem~\cite{Dornheim_JCP_2023,quantum_theory}
\begin{eqnarray}
  \frac{F_{\mathbf{q},A}-F_0}{N} =  \frac{  \chi(\mathbf{q},0) A^2}{n_0}\ ,
\end{eqnarray}
where the external potential in Eqs.~(1) of the main text is given by 
\begin{eqnarray}\label{eq:harmonic}
    v_\textnormal{ext}(\mathbf{r}) = 2A\ \textnormal{cos}\left(
\mathbf{r}\cdot\mathbf{q}
    \right)\ ;
\end{eqnarray}
for simplicity, we consider the ideal (i.e., noninteracting) Fermi gas, i.e., $\hat{W}\equiv0$.
The static density response function is readily computed from the imaginary-time version of the fluctuation--dissipation theorem~\cite{Dornheim_MRE_2023}
\begin{eqnarray}\label{eq:static_chi}
    \chi(\mathbf{q},0) = - n \int_0^\beta \textnormal{d}\tau\ F(\mathbf{q},\tau)\ ,
\end{eqnarray}
where the imaginary-time density--density correlation function $F(\mathbf{q},\tau)$ is computed from a PIMC simulation of the unperturbed ideal Fermi gas.

The corresponding extended ensemble is given by 
\begin{eqnarray}
    Z_\textnormal{extended} = Z_{\mathbf{q},A} + Z_0\ ,
\end{eqnarray}
where $Z_0$ is given by the ideal partition function and $Z_\mathbf{q,A}$ the ideal analogue within the external potential Eq.~(\ref{eq:harmonic}); we set $c_\eta\equiv 1$ as there is sufficient overlap between the perturbed and unperturbed configuration spaces.
In the linear response limit, the (fermionic) ratio of the two sectors is given by
\begin{eqnarray}\label{eq:stiff_ratio}
\frac{Z_{q,A}}{Z_0} =
    \frac{\braket{\delta_{q,A}\ S}'}{\braket{\delta_0\ S}'} = 
    e^{-\beta(F_{\mathbf{q},A}-F_0)}\ ,
\end{eqnarray}
and the corresponding estimate for bosons follow from setting $S\equiv1$ in Eq.~(\ref{eq:stiff_ratio}). PIMC results are compared with Eq.~(\ref{eq:stiff_ratio}) in Fig.~2 of the main text.



\subsection{Derivation of stiffness theorem for single external perturbation in periodic boundary conditions}\label{s:app1}

The stiffness theorem for the free energy reads \cite{moldabekov2024density}:
\begin{equation}\label{eq_app1:F_1}
    \Delta F[n]= \frac{1}{2} \int \left.\chi(\vec k)\left| v_{\rm ext}(\vec k)\right|^{2}\right. ~\mathrm{d}\vec k,
\end{equation}
where $\Delta F[n]$ is the change in the free energy due to the external perturbation Eq.~(\ref{eq:harmonic}), and we use the notation $\chi(\mathbf{k})\equiv\chi(\mathbf{k},0)$ for brevity.

For the present purposes, we need the representation using the discrete Fourier transform since we perform calculations in a cubic cell with periodic boundary conditions. Accordingly, for Eq. (\ref{eq_app1:F_1}) we have:
\begin{equation}\label{eq_app1:F_2}
    \Delta F[n]= \frac{1}{2\Omega} \sum_{\vec k} \left.\chi(\vec k)\left|v_{\rm ext}(\vec k)\right|^{2}\right.,
\end{equation}
where $\Omega=L^3$ is the volume of the simulation cell, and $\vec k$ attains discrete values commensurate with the simulation cell.

The discrete Fourier transform of potential (\ref{eq:harmonic}) yields
\begin{eqnarray}\label{eq_app1:vext}
    v_{\rm ext}(\vec k)=A\Omega \left[ \delta_{\vec k, \vec q}+\delta_{\vec k, -\vec q}\right]. 
\end{eqnarray}

Substituting Eq. (\ref{eq_app1:vext}) into Eq. (\ref{eq_app1:F_2}), we find:
\begin{equation}\label{eq_app1:F_3}
    \Delta F[n]= \frac{A^2 \Omega}{2} \left[ \chi(\vec q)+\chi(-\vec q)\right]. 
\end{equation}

From Eq. (\ref{eq_app1:F_3}), taking into account that $\chi(\vec q)=\chi(-\vec q)$ and  dividing $\Delta F[n]$ by the total number of particles $N$, we derive:
\begin{equation}\label{eq_app1:F_4}
    \frac{\Delta F[n]}{N}= \frac{A^2}{n_0}\chi(\vec q),
\end{equation}
where $n_0=N/\Omega$ and $\Delta F[n]=F_{\mathbf{q},A}-F_0$ is the change in the free energy due to perturbation (\ref{eq:harmonic}).

\section{Finite-size correction for uniform electron gas}

The finite-size error of the interaction energy is pre-dominantly given by the approximation of a continuous integral over the static structure factor $S(\mathbf{q})$ by a sum over discrete lattice vectors in the finite simulation box~\cite{Chiesa_PRL_2006,Drummond_PRB_2008,Holzmann_PRB_2016,dornheim_prl,review}. It can be accurately estimated via
    \begin{eqnarray}
        \Delta W(N) &=& \frac{1}{2} \int_{q<\infty} \frac{\textnormal{d}\mathbf{q}}{(2\pi)^3} \left\{\left( S(\mathbf{q})-1 \right) \frac{4\pi}{q^2}\right\} \\\nonumber & &
        - \left\{\frac{1}{2\Omega} \sum_{\mathbf{G}\neq\mathbf{0}}\left( S(\mathbf{G}) - 1\right) \frac{4\pi}{G^2} + \xi_\textnormal{M}\right\}\ ,
    \end{eqnarray}
where $\xi_\textnormal{M}$ is the Madelung constant~\cite{Fraser_PRB_1996}, and $S(\mathbf{q})$ can be evaluated e.g.~in the random phase approximation (RPA).

The corresponding finite-size correction of $f_\textnormal{xc}$ at a density parameter $r_s$ and reduced temperature $\Theta$ is then computed via~\cite{review} 
\begin{eqnarray}
  \Delta f_\textnormal{xc}(N) =   \frac{1}{r_s^2} \int_0^{r_s}\textnormal{d}\overline{r}_s\ \overline{r}_s \frac{\Delta W(\overline{r}_s,\Theta)}{N}\ ,
\end{eqnarray}
using the implementation in the \texttt{uegpy} code~\cite{uegpy} by F.D.~Malone.

\section{Free energy of hydrogen snapshots}

To compare the free energy of a hydrogen snapshot between different simulation methods, it is important to consistently define a reference energy level with respect to the ions. In the \texttt{ISHTAR} code~\cite{ISHTAR}, we follow the notation by Fraser \emph{et al.}~\cite{Fraser_PRB_1996} and define the Ewald potential $\psi(\mathbf{r}_1,\mathbf{r}_2)$ in a way that it averages to zero over the simulation cell.
The full snapshot Hamiltonian for a fixed set of proton positions $\mathbf{I}_k$ is given by
\onecolumngrid
\begin{eqnarray}\label{eq:full_Hamiltonian}
    \hat{H}_\textnormal{snap} = \underbrace{- \frac{\hbar^2}{2m_e}\sum_{l=1}^N \nabla_l^2}_{\hat{K}} + \underbrace{e^2\sum_{l>k}^N\psi(\hat{\mathbf{r}}_l,\hat{\mathbf{r}}_k)+ \frac{N}{2}e^2\xi_\textnormal{M}}_{\hat{W}_{ee}}
    + \underbrace{e^2 \sum_{l<k}^N\psi(\mathbf{I}_l,\mathbf{I}_k) - e^2\sum_{l=1}^N \sum_{k=1}^N \psi(\hat{\mathbf{r}}_k,\mathbf{I}_l) + \frac{N}{2}e^2\xi_\textnormal{M}}_{\hat{v}_\textnormal{ext}}\ ;\quad
\end{eqnarray}
\twocolumngrid
here $\xi_\textnormal{M}$ denotes the usual Madelung constant. The fist two grouped terms in Eq.~(\ref{eq:full_Hamiltonian}) denote the kinetic [$\hat{K}$] and interaction [$\hat{W}_{ee}$] energy of the electrons, respectively.
The final contribution is the external energy [$\hat{v}_\textnormal{{ext}}$] due to the fixed protons, which explicitly includes the electrostatic energy due to the proton--proton interaction (and the corresponding Madelung term). While the protonic contribution is constant for any given simulation, it does shift the value of the free energy $F$. In practice, it is included in both PIMC and DFT simulations~\cite{kresse1996efficient}, which allows us to compare the respective free energies.

\section{Computational details}

\subsection{Path integral Monte Carlo}

All PIMC simulations presented in this work are carried out using the extended canonical ensemble approach~\cite{Dornheim_PRB_nk_2021} implemented in the \texttt{Ishtar} code~\cite{ISHTAR}.
For the simulation of the harmonically perturbed system (Fig.~2 of the main text) and of the warm dense electron gas (Fig.~3 of the main text), we have used $P=100$ imaginary-time factors in the primitive approximation, which is sufficient for convergence within the available Monte-Carlo error bars.

For the simulation of the inhomogeneous ion snapshot, we have used $P=100$ imaginary-time factors using the pair approximation based on the approach described in Ref.~\cite{Bohme_PRE_2023}.

\subsection{Density functional theory}

The KS-DFT simulations were performed using the GPAW code~\cite{GPAW1, GPAW2, ase-paper, ase-paper2}, which is a real-space implementation of the projector augmented-wave (PAW) method~\cite{BlochlPAW}.
The energy cutoff is set to $800~{\rm eV}$.
We used the ground-state LDA XC functional by Perdew and Wang \cite{Perdew_Wang} and finite temperature LDA by Groth {et al.} \cite{groth_prl}.
PAW datasets automatically generated by GPAW have been used. 
The simulations are performed for $r_s=2$ and $\theta=2$. 
We used $N_b=680$ bands for $N=14$, $N_b=980$ for $N=20$, $N_b=1100$ for $N=32$, and $N_b=2100$ for $N=66$. The used number of bands ensures a smallest occupation number of about $10^{-5}$.
We set the $k$-point grid to $10\times10\times10$ for $N=14$ and $N=20$ particles, to $8\times8\times8$ for $N=32$ particles, and  to $4\times4\times4$ for $N=66$ particles.


%